\documentclass[fleqn,12pt,twoside]{article}
\usepackage{fleqn,espcrc1,epsfig,citesort}

\usepackage{graphicx}
\usepackage[figuresright]{rotating}

\newcommand{\mc}{\multicolumn}
\newcommand{\lsim}{\mathrel{\mathop{\kern 0pt \rlap
  {\raise.2ex\hbox{$<$}}}
  \lower.9ex\hbox{\kern-.190em $\sim$}}}
\newcommand{\gsim}{\mathrel{\mathop{\kern 0pt \rlap
  {\raise.2ex\hbox{$>$}}}
  \lower.9ex\hbox{\kern-.190em $\sim$}}}

\title{Non--Mesonic Weak Decay of $\Lambda$--hypernuclei:
      a new determination of the $\Gamma_n/\Gamma_p$ ratio}

\author{\underline{G. Garbarino}\address[to]{Dipartimento di Fisica Teorica,
        Universit\`a di Torino, and INFN, Sezione di Torino, I--10125 Torino, 
        Italy}\address[est]{Departament
        d'Estructura i Constituents de la Mat\`eria,
        Universitat de Barcelona, E--08028 Barcelona, Spain},   
        A. Parre\~{n}o\addressmark \, and A. Ramos\addressmark \thanks{Work 
        partly supported by EURIDICE HPRN--CT--2002--00311,
        DGICYT BFM2002--01868, Generalitat de Catalunya SGR2001--64 and
        INFN.}}

\begin{document}

\maketitle

\begin{abstract}

Theoretical descriptions of the non--mesonic weak decay of $\Lambda$--hypernuclei are unable 
to reproduce the experimental
values of the ratio $\Gamma_n/\Gamma_p\equiv \Gamma(\Lambda n\to nn)/
\Gamma(\Lambda p\to np)$. In this contribution we discuss a new
approach to this problem. We have incorporated a
one--meson--exchange model for the $\Lambda N\to nN$ transition in finite nuclei
in an intranuclear cascade code for the calculation
of double--coincidence nucleon distributions corresponding to
the non--mesonic decay of $^5_\Lambda {\rm He}$ and $^{12}_\Lambda {\rm C}$.
The two--nucleon induced decay mechanism, $\Lambda np\to nnp$, has been taken
into account within a local density approximation scheme using
a one--pion--exchange model supplemented by short range correlations. 
A weak decay model independent analysis of preliminary
KEK coincidence data for $^5_\Lambda {\rm He}$
allows us to extract $\Gamma_n/\Gamma_p=0.39\pm 0.11$ 
when the two--nucleon induced channel is neglected (i.e., $\Gamma_2=0$)
and $\Gamma_n/\Gamma_p=0.26\pm 0.11$ when $\Gamma_2/\Gamma_1=0.2$. 

\end{abstract}

\section{INTRODUCTION}
\label{intro}

An old challenge of hypernuclear studies has been to secure the
``elusive'' theoretical explanation of the large experimental 
values ($\simeq 1$) of the ratio, $\Gamma_n/\Gamma_p$, between the neutron-- and 
proton--induced non--mesonic 
decay rates, $\Gamma_n\equiv \Gamma(\Lambda n\to nn)$ and 
$\Gamma_p\equiv \Gamma(\Lambda p\to np)$ \cite{Al02,Ra98}.   

Because of its strong tensor 
component, the one--pion--exchange (OPE) model
supplies very small $\Gamma_n/\Gamma_p$ ratios, typically in the interval $0.05\div 0.20$
for $s$-- and $p$--shell hypernuclei. 
On the contrary, the OPE description can reproduce the total 
non--mesonic decay rates observed for these systems.
Other interaction mechanisms are then expected to correct for the 
overestimation of $\Gamma_p$ and the underestimation of $\Gamma_n$
characteristic of the OPE. 
Those which have been studied extensively in the literature are the 
following ones: i) the inclusion in the ${\Lambda}N\rightarrow nN$
transition potential of mesons heavier than the pion (also including the exchange
of correlated or uncorrelated two--pions)
\cite{Pa97,Os01,Pa01,It98}; ii) the inclusion of interaction terms that
explicitly violate the ${\Delta}I=1/2$ rule
\cite{Al02,Pa98,Al99b}; iii) the inclusion of the two--body induced decay mechanism
\cite{Al91,Ra95,Al00,Al99a} and iv) the description of the
short range $\Lambda N\to nN$ transition in terms of quark degrees of freedom
\cite{Ok99}, which automatically introduces $\Delta I=3/2$ contributions.    

Some progress in the theory of non--mesonic decay has been experienced 
in the last years. A few calculations \cite{Os01,Pa01,It98,Ok99}
with $\Lambda N \rightarrow nN$ transition potentials including
heavy--meson--exchange and/or direct quark contributions
obtained ratios more in agreement with data, without providing, nevertheless, 
an explanation of the origin of the puzzle \cite{Al02}.
Very recently, the $\Lambda N\to nN$ interaction has been studied within an 
effective field theory framework \cite{Pa03} with a weak decay model
consisting of OPE, one--kaon--exchange and 
$|\Delta S|=1$ four--fermion contact terms. 

In the light of the experiments under way and/or planned
at KEK \cite{Ou00a}, FINUDA \cite{FI01} and BNL \cite{Gi01},
it is important to develop different theoretical
approaches and strategies for the determination of $\Gamma_n/\Gamma_p$
from data. In this contribution we discuss an evaluation of nucleon--nucleon
coincidence distributions in the non--mesonic weak decay of $^5_\Lambda {\rm He}$ 
and $^{12}_\Lambda$C \cite{previous}. This work is motivated by the fact
that, in principle, correlation observables permit a {\it cleaner}
extraction of $\Gamma_n/\Gamma_p$ from data than single--nucleon observables.   
This is due to the elimination of interference terms
between $n$-- and $p$--induced decays \cite{Al02}, which are
unavoidable in experimental data and
cannot be taken into account by the Monte Carlo methods
usually employed to simulate the nucleon propagation through the residual nucleus. 
For a detailed discussion of this issue see Ref.~\cite{previous}.

The calculations are performed by combining a one--meson--exchange (OME)
model describing one--nucleon induced weak decays in finite nuclei 
with an intranuclear cascade code taking into account the nucleon final state 
interactions. The two--nucleon induced channel is also taken into account,
treating the nuclear finite size effects by means of a local density approximation 
scheme.

We also perform a weak interaction model independent analysis to extract an estimate
for $\Gamma_n/\Gamma_p$ using preliminary results from KEK \cite{Ou00a,OuOk}
on two--nucleon angular and energy correlations.
The resulting $\Gamma_n/\Gamma_p$ values for $^5_\Lambda {\rm He}$
turn out to be substantially smaller than those obtained
from single nucleon distributions analyses \cite{Sz91,No95a}
and fall within the predictions 
of recent theoretical studies \cite{Pa01,It98,Ok99}. 

The work is organized as follows. In Sec.~\ref{models} we give an
outline of the models employed to describe the non--mesonic weak decay
and we discuss the main features of the
intranuclear cascade simulation accounting for the nucleon propagation inside 
the residual nucleus. A selection of our results 
is discussed in Sec.~\ref{res} and the conclusions are given
in Sec.~\ref{conc}. 

\section{MODELS} 
\label{models}
\subsection{Weak decay}

The one--nucleon induced non--mesonic decay rates and the distributions of the 
nucleons produced in these processes are obtained with the OME model
of Refs.~\cite{Pa97,Pa01}. 
The OME weak transition potential 
contains the exchange of $\rho$, $K$, $K^*$, $\omega$ and 
$\eta$ mesons in addition to the pion. 
The strong couplings and strong final state interactions acting
between the weak decay nucleons are taken into account by using a scattering $nN$ 
wave function from the Lippmann--Schwinger ($T$--matrix) equation obtained with 
NSC97 (versions ``a'' and ``f'') potentials \cite{nsc}. 
The corresponding decay rates for $^5_\Lambda {\rm He}$ and $^{12}_\Lambda {\rm C}$
are listed in Table~\ref{gammas} (OMEa and OMEf)
together with the OPE predictions.
\begin{table}
\begin{center}
\caption{Weak decay rates (in units of the free $\Lambda$ decay width)
predicted by Ref.~\protect\cite{Pa01}
for $^5_\Lambda$He and $^{12}_\Lambda$C.}
\vspace{3mm}
\label{gammas}
\begin{tabular}{c|c c c|c c c}
\hline
\mc {1}{c|}{} &
\mc {1}{c}{} &
\mc {1}{c}{$\Gamma_n+\Gamma_p$} &
\mc {1}{c|}{} &
\mc {1}{c}{} &
\mc {1}{c}{$\Gamma_n/\Gamma_p$} &
\mc {1}{c}{} \\
                  &   OPE  &  OMEa  &  OMEf &   OPE  & OMEa   & OMEf   \\ \hline
$^5_\Lambda$He    & $0.43$ & $0.43$ & $0.32$ & $0.09$ & $0.34$ & $0.46$   \\
$^{12}_\Lambda$C  & $0.75$ & $0.73$ & $0.55$ & $0.08$ & $0.29$ & $0.34$   \\
\hline
\end{tabular}
\end{center}
\end{table}

The differential and total decay rates for the two--nucleon induced process, 
$\Lambda np\to nnp$, are calculated with the 
polarization propagator method in local density approximation
(LDA) of Refs.~\cite{Ra95,Al00}. 
In such a calculation, the simple OPE mechanism,
supplemented by strong $\Lambda N$ and $NN$ short range correlations (given 
in terms of phenomenological Landau functions), describes the weak 
transition process. In the present calculation,
the distributions of the nucleons emitted by two--nucleon stimulated
decays and the 
value of $\Gamma_2$ are properly scaled to maintain the ratio
$\Gamma_2/\Gamma_1$ unchanged: we then use 
$\Gamma_2/\Gamma_1\equiv (\Gamma_2/\Gamma_1)^{\rm LDA}=0.20$ for $^5_\Lambda$He and
$0.25$ for $^{12}_\Lambda$C.

\subsection{Intranuclear cascade simulation}
\label{mc}

In their way out of the nucleus, the weak decay (i.e., primary) nucleons
continuously change energy, direction and charge due to collisions with other nucleons. 
As a consequence, secondary nucleons are also emitted. 

We simulate the nucleon propagation inside the
residual nucleus with the Monte Carlo code of Ref.~\cite{Ra97}.
A random number generator determines the decay channel, $n$--, $p$--
or two--nucleon induced, according to the values of $\Gamma_n/\Gamma_p$
and $\Gamma_2/\Gamma_1$ predicted by our finite nucleus and LDA
approaches. Positions, momenta 
and charges of the weak decay nucleons are selected by the same
random number generator, according to the corresponding
probability distributions given by the finite nucleus and LDA 
calculations. 

Once they are produced, the primary nucleons move under a local potential
$V_N(R)=-k_{F_N}^2(R)/2 m_N$, where 
$k_{F_N}(R)=[3\pi^2\rho_N(R)]^{1/3}$ ($N=n,p$) is the
local nucleon Fermi momentum corresponding to the nucleon density $\rho_N(R)$. 
The primary nucleons also collide
with other nucleons of the medium according to free space
nucleon--nucleon cross sections \cite{Cu96} properly corrected 
to take into account the Pauli blocking effect. For further
details concerning the intranuclear cascade calculation see 
Ref.~\cite{Ra97}. Each Monte Carlo event ends
with a certain number of nucleons which leave the nucleus
along defined directions and with defined energies. One can then 
select the outgoing nucleons and store them in different ways,
as we shall do in the discussion of Section~\ref{res}.

\section{RESULTS}
\label{res}

The ratio $\Gamma_n/\Gamma_p$ is defined as the ratio
between the number of weak decay $nn$ and $np$ pairs,
$N^{\rm wd}_{nn}$ and $N^{\rm wd}_{np}$. However, due to 
two--body induced decays and (especially) nucleon FSI effects, one expects
the following inequality:
\begin{equation}
\label{ratio-nn}
\frac{\Gamma_n}{\Gamma_p}\equiv \frac{N^{\rm wd}_{nn}}{N^{\rm wd}_{np}}
\neq \frac{N_{nn}}{N_{np}}\equiv
R_2\left[\Delta \theta_{12}, \Delta T_n, \Delta T_p\right] ,
\end{equation}
when the observable numbers
$N_{nn}$ and $N_{np}$ are determined by employing particular intervals 
of variability of the pair opening angle, $\Delta \theta_{12}$, and the nucleon
kinetic energies, $\Delta T_n$ and $\Delta T_p$.
The results discussed in Ref.~\cite{previous} clearly show the dependence of $N_{nn}/N_{np}$
on $\Delta \theta_{12}$ and $\Delta T_n$ and $\Delta T_p$. However,
$N_{nn}/N_{np}$ turns out to be much less sensitive
to FSI effects and variations of the energy
cuts and angular restrictions than $N_{nn}$ and $N_{np}$ separately.    

The numbers of nucleon pairs $N_{NN}$ ---which we consider to be
normalized \emph{per non--mesonic weak decay}---
are related to the corresponding quantities for the neutron-- ($N^{\rm 1Bn}_{NN}$)
proton-- ($N^{\rm 1Bp}_{NN}$) and two--nucleon ($N^{\rm 2B}_{NN}$) induced processes
via the following equation:
\begin{equation}
\label{1-2}
N_{NN}=\frac{N^{\rm 1Bn}_{NN}\, \Gamma_n+ N^{\rm 1Bp}_{NN}\, \Gamma_p+N^{\rm 2B}_{NN}\, \Gamma_2}
{\Gamma_n +\Gamma_p+\Gamma_2}
\equiv N^{\Lambda n\to nn}_{NN}+ N^{\Lambda p\to np}_{NN}+
N^{\Lambda np\to nnp}_{NN} . \nonumber
\end{equation} 
Here, $N^{\rm 1Bn}_{NN}\equiv N^{\Lambda n\to nn}_{NN} (\Gamma_n +\Gamma_p+\Gamma_2)/\Gamma_n$ 
is normalized per neutron--induced non--mesonic weak decay, etc. 
Therefore, $N^{\rm 1Bn}_{NN}$, $N^{\rm 1Bp}_{NN}$ and $N^{\rm 2B}_{NN}$
($N^{\Lambda n\to nn}_{NN}$, $N^{\Lambda p\to np}_{NN}$ and $N^{\Lambda np\to nnp}_{NN}$) 
{\it do not} depend ({\it do} depend) on the interaction model
employed to describe the weak decay.

In Table \ref{ome-che} the ratio $N_{nn}/N_{np}$ predicted by the
OPE, OMEa and OMEf models for $^5_\Lambda$He 
and $^{12}_\Lambda$C is given for two opening angle intervals
and for $T_n, T_p\geq 30$ MeV. The results of the OMEa and OMEf
models are in reasonable agreement with the preliminary KEK--E462 data 
for $^5_\Lambda$He \cite{OuOk}.
\begin{table}
\begin{center}
\caption{Predictions for the ratio $R_2\equiv N_{nn}/N_{np}$
for $^5_\Lambda$He and $^{12}_\Lambda$C.
An energy threshold of $30$ MeV and two pair opening angle
regions have been considered. The (preliminary) data are from KEK--E462 \protect\cite{OuOk}.}
\vspace{3mm}
\label{ome-che}
\begin{tabular}{c|c c | c c}
\hline
\mc {1}{c|}{} &
\mc {1}{c}{$^5_\Lambda$He} &
\mc {1}{c|}{} &
\mc {1}{c}{$^{12}_\Lambda$C} &
\mc {1}{c}{} \\
       & cos $\theta_{NN}\leq -0.8$ & all $\theta_{NN}$ &
cos $\theta_{NN}\leq -0.8$ & all $\theta_{NN}$ \\ \hline
OPE     & $0.25$  & $0.26$  & $0.24$ & $0.29$  \\
OMEa    & $0.51$  & $0.45$  & $0.39$ & $0.37$  \\
OMEf    & $0.61$  & $0.54$  & $0.43$ & $0.39$ \\ \hline
EXP  & $0.44\pm 0.11$  &     &        &  \\
\hline
\end{tabular}
\end{center}
\end{table}   

\subsection{A weak interaction model independent analysis of data}

We now discuss a weak interaction model independent analysis
of KEK coincidence data. To this aim, we make use of the 6 quantities
$N^{\rm 1Bn}_{nn}$, $N^{\rm 1Bp}_{nn}$, $N^{\rm 2B}_{nn}$,
$N^{\rm 1Bn}_{np}$, $N^{\rm 1Bp}_{np}$ and $N^{\rm 2B}_{np}$ 
entering Eq.~(\ref{1-2}): they are quoted in Table~\ref{univ} and,
by definition, do not depend on the model used to describe the weak decay.
We have thus to employ the following relation:
\begin{equation}
\label{fit}
\frac{N_{nn}}{N_{np}}=\frac{\displaystyle\left(N^{\rm 1Bn}_{nn}+N^{\rm 2B}_{nn}
\frac{\Gamma_2}{\Gamma_1}\right) \frac{\Gamma_n}{\Gamma_p}
+N^{\rm 1Bp}_{nn}+N^{\rm 2B}_{nn}\frac{\Gamma_2}{\Gamma_1}}
{\displaystyle\left(N^{\rm 1Bn}_{np}+N^{\rm 2B}_{np} \frac{\Gamma_2}{\Gamma_1}\right)
\frac{\Gamma_n}{\Gamma_p}
+N^{\rm 1Bp}_{np}+N^{\rm 2B}_{np}\frac{\Gamma_2}{\Gamma_1}} , 
\end{equation}
using $\Gamma_n/\Gamma_p$ and $\Gamma_2/\Gamma_1$ as fitting parameters.
\begin{table}
\begin{center}
\caption{Predictions for the weak interaction model independent quantities
$N^{\rm 1Bn}_{nn}$, $N^{\rm 1Bp}_{nn}$, $N^{\rm 2B}_{nn}$,
$N^{\rm 1Bn}_{np}$, $N^{\rm 1Bp}_{np}$ and $N^{\rm 2B}_{np}$ 
(integrated over all angles and for
$T_n, T_p\geq 30$ MeV) of Eq.~(\ref{1-2})
for $^5_\Lambda$He and $^{12}_\Lambda$C. The numbers in parentheses
correspond to the angular region with cos~$\theta_{NN}\leq -0.8$.}
\vspace{3mm}
\label{univ}
\begin{tabular}{c|c c c}
\hline
\mc {1}{c|}{} &
\mc {1}{c}{$N^{\rm 1Bn}_{nn}$} &
\mc {1}{c}{$N^{\rm 1Bp}_{nn}$} &
\mc {1}{c}{$N^{\rm 2B}_{nn}$} \\ \hline
$^5_\Lambda$He    & $0.84$ ($0.53$) & $0.10$ ($0.02$) & $0.54$ ($0.34$)   \\
$^{12}_\Lambda$C  & $0.56$ ($0.30$) & $0.27$ ($0.05$) & $0.30$ ($0.12$)   \\ \hline
   & $N^{\rm 1Bn}_{np}$ & $N^{\rm 1Bp}_{np}$ & $N^{\rm 2B}_{np}$ \\ \hline
$^5_\Lambda$He    & $0.20$ ($0.05$) & $0.98$ ($0.49$) & $0.55$ ($0.22$)   \\
$^{12}_\Lambda$C  & $0.33$ ($0.08$) & $1.22$ ($0.38$) & $0.38$ ($0.11$)   \\
\hline
\end{tabular}
\end{center}
\end{table}     

\begin{figure}
\begin{center}
\mbox{\epsfig{file=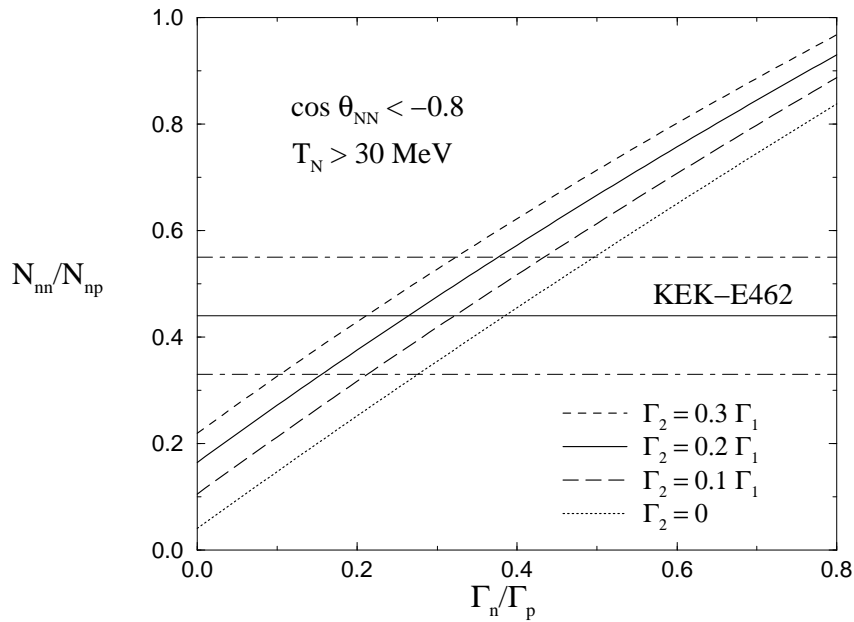,width=.68\textwidth}}
\caption{Dependence of the ratio $N_{nn}/N_{np}$ on $\Gamma_n/\Gamma_p$
and $\Gamma_2/\Gamma_1$ for $^5_\Lambda$He.
The results correspond to a nucleon energy threshold of $30$ MeV and
the angular restriction cos~$\theta_{NN}\leq -0.8$.
The horizontal lines show the
preliminary KEK--E462 data of Ref.~\protect\cite{OuOk}.}
\label{fithe}
\end{center}
\end{figure}

In Fig.~\ref{fithe} we report the dependence of $N_{nn}/N_{np}$ for
$^5_\Lambda$He on $\Gamma_n/\Gamma_p$ for four different 
values of $\Gamma_2/\Gamma_1$. The figure corresponds
to the experimentally interesting case with a nucleon energy 
threshold of $30$ MeV and the angular restriction
cos $\theta_{np}\leq -0.8$. For a given value of $\Gamma_2/\Gamma_1$,
Fig.~\ref{fithe} permits
an immediate determination of $\Gamma_n/\Gamma_p$ by a direct comparison with
data for the observable $N_{nn}/N_{np}$.  

By using the $^5_\Lambda {\rm He}$ data $N_{nn}/N_{np}=0.44\pm0.11$ 
from KEK--E462 \cite{OuOk} and neglecting the two--nucleon induced mechanism
(i.e., $\Gamma_2=0$), Eq.~(\ref{fit}) supplies: 
\begin{equation} 
\label{fit1}
\frac{\Gamma_n}{\Gamma_p}\left(^5_\Lambda {\rm He}\right)=0.39\pm0.11 . 
\end{equation} 
By employing the value $\Gamma_2/\Gamma_1=0.2$ (i.e., the one
obtained with the model of Ref.~\cite{Al00} and used in the present calculation), 
a 34\% reduction of the ratio is predicted: 
\begin{equation} 
\label{fit2}
\frac{\Gamma_n}{\Gamma_p}\left(^5_\Lambda {\rm He}\right)=0.26\pm 0.11 . 
\end{equation} 
These results for $\Gamma_n/\Gamma_p$ are in agreement with the pure theoretical
predictions of Refs.~\cite{Pa01,It98,Ok99}. On the contrary, they
are rather small if compared with previous determinations \cite{Sz91} ($0.93\pm 0.55$) 
\cite{No95a} ($1.97\pm 0.67$) from single--nucleon spectra analyses.  
Actually, all the previous experimental analyses of
single--nucleon spectra \cite{exp,Sz91,No95a}, supplemented in some cases 
by intranuclear cascade calculations, derived $\Gamma_n/\Gamma_p$ values in 
disagreement with pure theoretical predictions. In our
opinion \cite{previous}, the fact that our calculations reproduce coincidence data 
for values of $\Gamma_n/\Gamma_p$ as small as $0.3\div 0.4$ could signal 
the existence of non--negligible interference effects between the $n$-- and
$p$--induced channels in those old single--nucleon data.

Forthcoming coincidence data from KEK and FINUDA
could be directly compared with the results presented here and in Ref.~\cite{previous}.
This will permit to achieve new determinations of $\Gamma_n/\Gamma_p$
---which will help in clarifying better the origin of the $\Gamma_n/\Gamma_p$ puzzle---
and to establish the first constraints on $\Gamma_2/\Gamma_1$.     

\section{CONCLUSIONS}
\label{conc}
To summarize,
our weak interaction models supplemented by FSI through an intranuclear
cascade simulation provide double--coincidence observables which
are in reasonable agreement with preliminary KEK--E462 data for $^5_\Lambda$He. 

Through a weak interaction model independent analysis in which  
$\Gamma_n/\Gamma_p$ and $\Gamma_2/\Gamma_1$ are free parameters we
reproduce the KEK $^5_\Lambda {\rm He}$ data $N_{nn}/N_{np}=0.44\pm 0.11$ if
$\Gamma_n/\Gamma_p=0.39\pm 0.11$ and $\Gamma_2=0$ 
or $\Gamma_n/\Gamma_p=0.26\pm 0.11$ and $\Gamma_2/\Gamma_1=0.2$.
Although these values of $\Gamma_n/\Gamma_p$ extracted from data
agree with other recent pure theoretical evaluations
(such an agreement has been achieved now for the first time), they are
rather small if compared with the results of previous analyses
from single--nucleon spectra. We suspect that non--negligible
interference effects between the neutron-- and proton--induced
channels affected those single--nucleon analyses.

In conclusion, although further (theoretical and experimental) work is 
needed, we think that our investigation proves how the study of nucleon
coincidence observables can offer a promising possibility
to solve the longstanding puzzle on the $\Gamma_n/\Gamma_p$ ratio.

\end{document}